\documentstyle[12pt]{article}
\begin{document}
\title{
\begin{flushright}
{\small SFU-HEP-110-93 }
\end{flushright}
\vspace{0.5cm}
Positive Definite Gravitational Action,
      \\
      Cosmological Constant
\\  and \\
Superstring Theory
}
\author{
 I.V.Volovich\thanks{Permanent address:  Steklov Mathematical
Institute, Vavilov st.42, GSP-1,117966, Moscow , Russia; E-mail:
volovich@mph.mian.su}\\
Department of Physics, Simon Fraser University\\
Burnaby, British Columbia, V5A 1S6, Canada}
\date { }
\maketitle
\begin{abstract}
We use the recently discovered universality of Einstein equations
in the first order formalism to suggest a positive
definite Euclidean
action. Possible implications for quantum gravity are considered.
We discuss the Hawking and Coleman approach to the  vanishing
of the  cosmological
constant by using the new action and find that the cosmological
constant is probably zero. A possible scenario for obtaining such
an action from superstring field theory is discussed.

\end{abstract}
\newpage

There is a well known problem in quantum gravity
that the euclidean gravitational action is not
bounded from below \cite{GHP}. This is unfortune
because it questions the elegant Hartle - Hawking
no boundary proposal for the wave function of the
universe \cite{HH}
and the Hawking and Coleman approach to the vanishing of
the cosmological constant \cite{Haw,Col}.

In this note we suggest a positive definite euclidean
gravitational action for Einstein equations
and use it to discuss the problem of the cosmological constant .

An explanation of the extreme smallness (or vanishing)
of the cosmological constant is one of the biggest challanges
in modern physics, for a review see \cite{Wei}.
In the context of superstring theory \cite{GSW}
we should explain why the cosmological constant
vanishes after supersymmetry breaking. So, let us
imagine that there is a fundamental theory, like
superstring theory, with the vanishing cosmological
constant and after supersymmetry breaking it
generates an effective gravitational action
$I[g,\Gamma ]$. We admit a dependence of the action
not only on the metric $g$ but also on the connection
$\Gamma$.  Recall that in supersymmetric theories
one uses the vielbein $e$ and
 the spin connection $\omega$ describing
an interaction with fermions.
As is known the $(g,\Gamma )$-formalism
is locally equivalent to the $(e,\omega)$-formalism.
 This first order ('Palatini')
formalism with an independent variation over
the metric
and the connection for the standard Hilbert
action leads to Einstein equations.

 What general conditions should  the action
 $I[g,\Gamma ]$ satisfy? It seems that there are at
 least three appropriate
 conditions on the action $I[g,\Gamma ]$:

- It should lead to  Einstein equations.

- Gravitational energy for an asymptotically
 flat space-time
should be positive.

- The euclidean version of the action should be
 bounded from below.

The Hilbert action does not satisfy  the last
condition. However actions satisfying all of the
 above conditions do exist.
The simplest one is
\begin{equation}
I[g,\Gamma ]=G^{2}\int (R^{2}+G^{2}\Lambda ^{2})^{2}\sqrt{g}d^4 x
\label{1}
\end{equation}
Here $R=R(g,\Gamma )$ is the scalar curvature depending on $g$
 and $\Gamma $, $R=g^{\mu \nu}R_{\mu \nu}(\Gamma )$, where
  $R_{\mu \nu}(\Gamma )$ is the Ricci tensor, $R_{\mu \nu}(\Gamma )=
R^{\sigma}_{\mu \sigma \nu}(\Gamma )$ with
$R^{\sigma}_{\mu \lambda \nu}(\Gamma )$ being the Riemannian tensor of
the symmetric connection $\Gamma ^{\lambda}_{\mu \nu}$, and $G$ and
$\Lambda $ are constants. Note that the action (\ref{1}) does not
contain a  term linear in $R$. An action with such a
 term we will
be discussed later.

The fact that the action (\ref{1}) leads to Einstein equations, but
not to a theory with higher derivatives as one would think, is a
manifestation of the recently discovered
universality of Einstein
equations \cite{FFV1}. This universality  means
 the following.
Consider an action
\begin{equation}
I[g,\Gamma ]=\int f(R) \sqrt{g}d^4x
\label {3}
\end{equation}
where $f(R)$ is an arbitrary analytic function
 of $R=R(g,\Gamma )$.
Independent variation of the action (\ref{3}) with
 respect to $g$ and
$\Gamma $ gives
\begin{equation}
f'(R)R_{(\mu \nu )}(\Gamma )~-~\frac{1}{2}f(R)g_{\mu \nu}=0
\label{4}
\end{equation}
\begin{equation}
\nabla _{\mu}(f'(R)\sqrt{g}g^{\lambda \sigma })=0
\label{5}
\end{equation}
Here $R_{(\mu \nu )}$ is the
symmetric part of $ R_{\mu \nu }$. Taking the trace
 of equation (\ref{4})
one gets
\begin{equation}
f'(R)R -2f(R)=0.
\label{6}
\end{equation}
Equation (5) is to be solved for $R$ given an
$f=f(R)$, which, in  general,  gives $R(g,\Gamma )=c$,
 where $c$ is
 a (constant) root of
equation (\ref{6}).
  Substituting this into (\ref{5})
under assumption
$$f'(c)\neq 0$$
one gets that the connection $\Gamma$ should be
the Levi-Civita connection
for the metric $g$. Then (\ref{4})
gives the Einstein equations
\begin{equation}
R_{\mu \nu }(g)=\frac{c}{4} g_{\mu \nu}
\label{7}
\end{equation}
where $R_{\mu \nu }(g)$ is the Ricci tensor of the metric $g$.

If $f'(c)=0$ or $f(R)=R^2$ (conformal invariant Lagrangian)
then one has a bifurcation point in the space of all
Lagrangians and one gets instead of the Einstein
equations more general equations.

If one has an interaction with matter fields with a traceless
energy-momentum tensor, one gets the same equation (\ref{6}).
This universality, i.e. independence of Einstein equations
on the Lagrangian, was used in 2-dimensional gravity \cite {FFV2}
and in string theory \cite {Vol1}.

For the action (\ref{1})  equation (\ref{6}) has two solutions
\begin{equation}
R=\pm G\Lambda
\label{9}
\end{equation}
and Einstein equations are
\begin{equation}
R_{\mu \nu }(g)=\pm \frac{G\Lambda }{4} g_{\mu \nu}.
\label{10}
\end{equation}

In the semiclassical approximation one can
 interpret $\exp (-I)$
as being proportional to the probability of
 the corresponding Euclidean
field configuration \cite{Haw}. Hawking  has
 considered the probability
for the Hilbert action and he has concluded that the most
probable configurations will be  those with
very small values of the
effective cosmological constant \cite{Haw}.
 Coleman \cite{Col}
 arrived at the same
conclusion arguing that wormholes have the effect
 of making the
cosmological constant vanish.  This mechanism was
 considered in
\cite{Rub,Gid,Pre,Hos}. In particular,
 in \cite{Vol2} it was noted that such a
mechanism in multidimensional space-time could
explain why one has to
consider a compactification procedure  instead of
 staying in
the underling ten-dimensional space-time on which
 superstrings propagate.
So,  Hawking-Coleman approach looks very
 attractive but it suffers
from the unboundedness of the standard Euclidean
action. (For a discussion
see \cite{Fis,Unr}).

Now, let us consider an application of the
 positive action (\ref{1}) to
the problem of the cosmological constant.
 For the 4-sphere of  radius $r$
one has $R_{\mu \nu }(g)=(3/r^{2})g_{\mu \nu}$ and
from (\ref{10}) one has
$r^{2}=12/G\Lambda$. The value of the
 action (\ref{1}) for this solution
is equal to
\begin{equation}
I=f_{0}\Omega _{4}r^{4},
\label{11}
\end{equation}
where $ f_{0}=4G^6 \Lambda^4$ and
 $\Omega _{4}=\frac{8}{3}\pi ^{2}$ is
 the volume of the
4-sphere of the
unit radius. Therefore one has
\begin{equation}
I=2^9 3\pi^2 (G^{2}\Lambda )^{2},
\label{12}
\end{equation}
The probability of such a configuration
will be of the order
\begin{equation}
\exp (-2^9 3\pi^2 (G^{2}\Lambda )^{2})
\label{14}
\end{equation}
Though  the function (\ref{14})  has no
 essential singularity in $\Lambda$
\cite{Haw,Col}, it is clear that the
most probable configurations will be those
 with very small values of
$\Lambda $.

Let us discuss the action (\ref{1}). It is invariant under
 the transformation
\begin{equation}
\Lambda  ~\to ~ -\Lambda
\label{14a}
\end{equation}
For a given constant $\Lambda $
in accordance with equation (\ref{9}) there are two
 disconnected sectors in
the space of solutions. One chooses one of them
 to live within. The
symmetry (\ref{14a})
remainds an antipode symmetry
 considered by Linde \cite{Lin}.
 To choose an appropriate sector one can
look to the positivity of the gravitational energy.
 The energy-momentum
complex for the action (2) was considerd
in \cite{BFFV} and it was
shown that it is
proportional to the Komar complex known for
the  Hilbert
action. The real time action $S$ corresponding to (\ref{1}) is
\begin{equation}
S[g,\Gamma ]=-G^{2}\int (R^{2}+G^{2}\Lambda ^{2})^{2}\sqrt{g}d^4 x
\label{15}
\end{equation}
and one can show that one of the two sectors
has a positive energy.

Because of the FFV-universality of Einstein
 equations \cite{FFV1}
one can use almost any arbitrary positive function
$f(R)$ to get a positive definite euclidean action.
For example, for the action

\begin{equation}
I[g,\Gamma ] =G^{2n-2}\int (R^2 +G^2 \Lambda^2)^n\sqrt g d^4x,
\label{2}
\end{equation}

where $n\not =1$, one has $R=\pm G\Lambda /\sqrt{n-1}$
and the value of the action on the sphere is

\begin{equation}
I=\frac{2^7 3\pi^2 n^n}{(n-1)^{n-1}}(G^2\Lambda)^{2n-2}.
\label{17}
\end{equation}

For a quadratic action
\begin{equation}
I[g,\Gamma ] =\frac{1}{16\pi G}\int (2 \Lambda -R+aR^2)\sqrt g d^4x,
\label{18}
\end{equation}

where the $R^2$-term can be considered for example as a quantum
correction to the standard Hilbert-Einstein action,
one has the Einstein equations

\begin{equation}
R_{\mu\nu}(g)=\Lambda g_{\mu\nu}
\label{19}
\end{equation}

and the value of the action on the sphere is
\begin{equation}
I=\frac{3\pi(8\Lambda a-1)}{G\Lambda}
\label{20}
\end{equation}
The point $~8\Lambda a=1~~$ is a bifurcation point.
 The probability is of the
order $~$ $\exp(-24\pi a/G)$ for large $\Lambda$
and $\exp(3\pi /G\Lambda)$
for small $\Lambda$. This important case
 deserves a further investigation.

A simple positive action with
a linear $R$-term is
\begin{equation}
I[g,\Gamma ] =\kappa^2\int (R^2 +aR+b)^2\sqrt g d^4x
\label{21}
\end{equation}
As is known $R(g,\Gamma )=R(e,\omega)$ where $R(e,\omega)=
R^{ab}_{\mu\nu}(\omega)e^{\mu}_{a}e^{\nu}_b$
and one can use an action
\begin{equation}
I[e,\omega]=\int f(R)\det e d^4x
\label{22}
\end{equation}
depending on the vielbein $e$ and the spin connection $\omega$
to derive Einstein equations.

The Hartle-Hawking wave function of the universe for the action
(\ref{22}) reads:
\begin{equation}
\Psi=\int e^{-I[e,\omega]}DeD\omega.
\label{23}
\end{equation}
It depends on boundary terms which we will consider in a separate
publication. Quantum gravity in the first order formalism
 has been
considered recently by Witten \cite{Wit1}. To study non-perturbative
quantum gravity  it would be important
to elaborate on a lattice formulation for the integral (\ref{23}) with
a positive action. An appropriate lattice formulation for a connection
has been suggested by Caselle, D'Adda and Magnea  \cite{CDM}.

Let us now discuss  what could be a scenarium leading to a positive
effective action, like (\ref{22}), from the superstring
theory. As we saw above
the problem of the cosmological constant is related with off-shell
formulation of the  theory. On shell we have Einstein equations but
off-shell there are many possible actions leading to the same
Einstein equations. In particular there is a class of actions  (\ref{3})
which is positive and for which the most probable value
of the cosmological constant is zero. How can one get
an action (\ref{3}) or (\ref{22}) from the superstring theory,
which is currently
the primary approach to unification of all physical laws?

The effective action for the metric resulting from integrating over
quantum fluctuations of matter is usually represented as an
expansion in powers of the curvature tensor and its derivatives:
\begin{equation}
S_{eff}[g]=\int d^{4}x\sqrt g (2\Lambda+aR+
bR_{\mu\nu\lambda\sigma}R^{\mu\nu\lambda\sigma}+...)
\label{24}
\end{equation}

If we approximate $S_{eff}[g]$ by Einstein gravity then
we take only the first two terms in the expansion (\ref{24})
and we get the standard action unbounded from below.
Note, however, that though the expansion (\ref{24}) takes place
for an arbitrary local functional of the metric,
it does not mean that such a functional will appear
in a superstring theory.
It seems that if we start with a reasonable fundamental
theory, like the superstring theory, and  use a reasonable
approximation we  will not get an effective action of type (\ref{24})
 with its dissapointing properties (unboundness)
but rather we would likely get an effective action like
(\ref{22}) which gives Einstein equations and still it is positive.
Let us recall that in the $\sigma$-model approach
to superstring theory from the vanishing of $\beta$-functions
 one gets only
equations of motion for target space fields but not an action.
As we have discussed there is a freedom in choosing an action
for the Einstein equations.
To get an action one needs an off-shell formulation
of superstring field theory.

As is known the presence of the spin connection in supergravity
is necessary to describe an interaction of gravity with
fermions \cite {New}. In modern approaches to superstring
theory one works on-shell, so one can not see a dependence
on the connection. Closed string field theory in modern
formulation does  not consider the  connection in the target space
as an independent variable \cite {Zwi}. However it seems
difficult to imagine a closed superstring field theory
containing only the metric on target space and
without the vielbein and the spin connection because one
 should reproduce
supergravity as an effective theory. In fact, a suggestion was made
\cite {AV} that a closed superstring field theory
should be formulated in terms of a vielbein
formalism including a spin connection on the loop space.
 Such a theory
probably could be a sort of topological supergravity on the loop space.
It seems
that the space of all two-dimensional fields \cite {Wit2}
has enough scope for such a theory. Such an approach,
if developed, could lead to an effective action
(\ref{22}) which gives Einstein equations and remains
positive and therefore
could be used for an explanation of the vanishing of the
cosmological constant after supersymmetry breaking.
$$~$$
$$~$$
{\bfAcknowledgments}
$$~$$
I am grateful to I.Ya.Aref'eva, M.Atiyah,
 A.Morosov, R.Myers, I.Tkachov,
K.S.Vis\-wanathan and W.G.Unruh for interest in this work
 and useful
discussions. This work has been supported
in part by an operating grant
of K.S.Viswanathan from Natural Sciences and
 Engeneering Research
Council of Canada. I am grateful to the Department of Physics
of the Simon Fraser University where this work was completed,
for the kind hospitality.

\end{document}